\documentclass{aa}
\usepackage[varg]{txfonts}
\usepackage[T1]{fontenc}
\usepackage[utf8]{inputenc}
\bibpunct{(}{)}{;}{a}{}{,}

\begin{document}

\title{An improved Magellan weak lensing analysis of the galaxy cluster Abell 2744}
\author{Davide Abriola\inst{1}\thanks{davide.abriola@unimi.it} \and Daniele Della Pergola\inst{1} \and Marco Lombardi\inst{1} \and Pietro Bergamini\inst{1,2} \and Mario Nonino\inst{3} \and Claudio Grillo\inst{1,4} \and Piero Rosati\inst{2,5}}

\institute{Dipartimento di Fisica, Università degli Studi di Milano, 
Via Celoria 16, I-20133 Milano, Italy \and INAF-OAS, Osservatorio di Astrofisica e Scienza dello Spazio di Bologna, via Gobetti 93/3, I-40129 Bologna, Italy \and INAF-OAT Osservatorio Astronomico di Trieste, via G. B. Tiepolo 11, I-34131 Trieste, Italy \and INAF-IASF Milano, via A. Corti 12, I-20133 Milano, Italy \and Dipartimento di Fisica e Scienze della Terra, Università degli Studi di Ferrara, via Saragat 1, I-44122 Ferrara, Italy}
\date{}

\abstract{We present a new weak lensing analysis of the Hubble Frontier Fields galaxy cluster Abell 2744 ($z$ = 0.308) using new Magellan/MegaCam multi-band $g$$r$$i$ imaging data. We carry out our study by applying brand-new PSF and shape measurement softwares that allow for the use of multi-band data simultaneously, which we first test on Subaru/Suprime-Cam $BR_cz'$ imaging data of the same cluster. The projected total mass of this system within $2.35 \, \mathrm{Mpc}$ from the south-west BCG is $(2.56 \pm 0.26) \times 10^{15} \, \mathrm{M}_\odot$, which makes Abell 2744 one of the most massive clusters known. This value is consistent, within the errors, with previous weak lensing and dynamical studies. Our analysis reveals the presence of three high-density substructures, thus supporting the picture of a complex merging scenario. This result is also confirmed by a comparison with a recent strong lensing study based on high-resolution JWST imaging. Moreover, our reconstructed total mass profile nicely agrees with an extrapolation of the strong lensing best-fit model up to several Mpc from the BCG centre.}


\maketitle

\section{Introduction}
\label{intro}

Clusters of galaxies are the largest gravitationally bound systems in the Universe, being composed of hundreds up to thousands of galaxies immersed in a diffuse halo of dark matter (DM), which constitutes up to $\sim 85 \%$ of their total mass \citep{david}. The remaining mass is subdivided between a hot and diffuse gas $-$ the Intra Cluster Medium (ICM), which amounts up to $\sim 15\%$ of the total mass $-$ and a stellar (galaxies) component. Galaxy clusters have been proved to be powerful cosmological probes. Indeed, being among the most massive systems of the Universe, they constitute the latest phase of the hierarchical structure formation. Moreover, they represent the ideal laboratory to study the evolution and interaction between galaxies. For several applications it is crucial to accurately reconstruct their total mass distribution \citep[see][]{pratt}. As a matter of fact, once the baryonic component has been mapped, one can infer the physical properties of DM and shed light on its nature \citep[e.g., see][]{clowe}. These findings can then be compared with the outcomes of cosmological simulations, thus to test the $\Lambda$CDM structure formation paradigm. 

To reach these aims, gravitational lensing is one of the most powerful tools, since, in contrast to other methods, it does not require any hypotheses regarding the physical nature, state equilibrium and mass composition of the deflector. Given their typical total masses ($\simeq 10^{13}-10^{15} \, \mathrm{M_\odot}$), the dense inner cores of galaxy clusters act as strong gravitational lenses, producing elongated arcs and hundreds of multiple images of background sources, many of which would otherwise not be observed. This has led to the discovery of galaxies lying at redshift higher than 10 \citep[e.g.,][]{atek2, atek1, borsani}. Therefore, these systems have been the target of several dedicated imaging and spectroscopic surveys, e.g. the Cluster Lensing And Supernova survey with Hubble \citep[CLASH,][]{clash}, its extension CLASH-VLT \citep{clash-vlt} based on deep (up to redshift 7) spectoscopic data acquired with the the spectrograph VIMOS \citep{vimos} at the Very Large Telescope (VLT), and the Hubble Frontier Fields campaign \citep[HFF,][]{lotz2017}. These studies have led to detailed and accurate analysis of the inner clusters cores, up to hundreds of $\mathrm{kpc}$ \citep[e.g.,][]{grillo, bergamini2019}, by exploiting the information arising from hundreds of spectroscopically-confirmed multiple images. 

On the other hand, in the less dense regions of galaxy clusters, background galaxies are only weakly distorted. This is the weak lensing (WL) regime, which provides a complementary and independent probe to measure the total mass of galaxy clusters \citep[see, e.g.,][]{h2000, lombardi2, umetsu2010, umetsu2014, umetsu2022}. Indeed, the statistical study of the slight distortion induced on samples of background galaxies allows for a robust and model-free \citep{kaiser} reconstruction of the total mass distribution up to the outskirts of these systems (i.e., up to few Mpc from their centres), where no multiple images are produced. Hence, combined strong and weak lensing studies allow one to map the projected total mass of the deflectors on different scales and a good agreement between the two probes has been observed in several clusters \citep[e.g.,][]{umetsu2011, coe, mede2013, niemiec}. \\

Among the HFF clusters, Abell 2744 \citep[lying at redshift $z_d = 0.308$,][]{lotz} shows one of the most complex merger phenomena ever detected, observed both in the radio \citep{radio1, radio2, radio3} and in the X-ray \citep{x1, x2} data. Therefore, this system has been the focus of several studies which revealed a north-south merger, including a lensing analysis by \citet{merten}, based on Hubble Space Telescope, Subaru and VLT imaging data, that suggested a complicated merging scenario, with the main cluster potential in the southern part. A recent WL analysis by \citet{mede2016} using Subaru/Suprime-Cam imaging supported the picture of multiple mergers, by discovering the presence of four substructures. This scenario has then been further explored by \citet{jau}, who performed a joint optical (using data from the HFF campaign) \& X-ray (using data collected by XMM-Newton) strong and weak gravitational lensing study of this cluster. Their analysis concluded that Abell 2744 is indeed a rare case of an extreme system. More recently, this system has been the target of still ongoing surveys, including the Beyond Ultra-deep Frontier Fields And Legacy Observations \citep[BUFFALO,][]{buffalo} survey, aimed at studying early galactic assembly and clustering, as well as the James Webb Space Telescope (JWST) Early Release Science (ERS) program Grism Lens Amplified Survey from Space \citep[GLASS,][]{treu}, the JWST Ultradeep NIRspec and NIRcam observations before the epoch of reionization \citep[UNCOVER, ][]{uncover} and the Director's Discretionary Time (DDT) observations program 2756 \citep[PI: Wenlei Chen,][]{chen2, chen1}. The high-resolution imaging data acquired by JWST have led to recent gravitational lensing analysis, including the strong lensing (SL) only study by \cite{bergamini2023b} and the combined free-form (non-parametric) SL\&WL one by \cite{cha}.\\

In this paper we present an improved WL analysis of the galaxy cluster Abell 2744 using deep Magellan/MegaCam multi-band imaging data covering a field of view of approximately $31' \times 33'$. We verify possible sources of systematic uncertainties, thus obtaining a robust WL total mass reconstruction which does not suffer from the dilution due to foreground sources, and compare our results with those of a complementary SL study by \cite{bergamini2023b}.

The paper is organised as follows. In Section \ref{methodology} we briefly recall the principles of WL, with the basic relations required to reconstruct the total mass distribution of the cluster starting from the shape measurements. In Section \ref{Data} we present the Magellan data, the detection of the astronomical sources and their classification between stars, cluster members and foreground galaxies. In Section \ref{analysis} we describe the WL procedure to obtain the total mass distribution of the cluster. The results obtained are discussed in Section \ref{discussion}, where we compare them with those from previous strong and weak lensing analyses. Finally, we summarise our findings in Section \ref{conclusions}.

Throughout this paper, we use the AB magnitude system and adopt a flat $\mathrm{\Lambda CDM}$ cosmology with $\Omega_m = 0.31$, $\Omega_\Lambda = 0.69$, and $H_0 = 67.66 \, \mathrm{km\, s^{-1} \, Mpc^{-1}}$. In this cosmology, $1'$ corresponds to $286 \,  \mathrm{kpc}$ at the cluster redshift, $z_d = 0.308$. The adopted cluster centre is $\mathrm{R. A.} = 3.58^\circ$, $\mathrm{Decl.} = - 30.4^\circ$ (J2000.0).

\section{WL methodology}
\label{methodology}

\noindent WL allows for the reconstruction of the dimensionless surface mass distribution (or convergence) $\kappa$ of a galaxy cluster starting from the measurements of the lensed shapes of background galaxies (i.e., lying at a redshift higher than that of the lens), expressed in terms of their complex ellipticity $\varepsilon = \varepsilon_1 + i \varepsilon_2$ \citep[for a review see, e.g.,][]{bartelmann}. $\kappa$ is related to the dimensional surface mass distribution $\Sigma$ through
\begin{equation}
    \kappa = \frac{\Sigma}{\Sigma_\mathrm{cr}} \, ,\end{equation}
where 
\begin{equation}
    \Sigma_\mathrm{cr} = \frac{c^2}{4 \pi \, G}\frac{D_\mathrm{s}}{D_\mathrm{d} \, D_\mathrm{ds}} \, 
\label{crit-sigma}
\end{equation}
is the critical surface mass density, a geometrical term defined in terms of the angular-diameter distances between the observer and the lens ($D_\mathrm{d}$), the observer and the source ($D_\mathrm{s}$), and the lens and the source ($D_\mathrm{ds}$).\\

\noindent Given the measurements of the shapes of background galaxies, we first obtain the two-dimensional shear field $\gamma = \gamma_1 + i \gamma_2$ induced by the gravitational potential of the deflector through a moving average over a pixelised grid. This can be done by applying the iterative method outlined by \citet{seitz}: we first start from a first-guess null dimensionless surface mass distribution $\kappa_0$, and estimate $\gamma$ using Eq. (\ref{shear-field}). We then use this value to recover $\kappa$ through Eq. (\ref{action}), and exploit this quantity to further estimate $\gamma$. This process is performed several times, until convergence is reached, which takes place within few steps. Since the mutual distance between the lens and the source affects the resulting map, we take into account the redshift distribution of the sources, thus evaluating the shear field $\gamma_\ell(\vec{x})$ at each position $\vec{x}$ of the grid and at the $\ell$-th iteration as
\begin{equation}
    \gamma_\ell(\vec{x}) = \frac{\sum_n K(\vec{x}, \vec{x}_n) \, \omega_n \, \left(1 - \frac{\langle Z \rangle_n}{\langle Z^2 \rangle_n}  \, \kappa_{\ell-1} \right) \varepsilon_n}{\sum_n K(\vec{x}, \vec{x}_n) \, \omega_n \, \langle Z \rangle_n^2} \, ,
\label{shear-field}
\end{equation}
where $K(\vec{x}, \vec{x}_n)$ is a kernel depending on the projected distance from the position $\vec{x}_n$ of the $n$th background galaxy and $\varepsilon_n$ its ellipticity. $\kappa_{\ell-1}$ is the dimensionless surface mass distribution estimated at the $(\ell-1)$-th iteration for a fictitious source lying at redshift infinite. The quantity $\omega_n$ is a weight factor that depends on the uncertainty $\sigma_n$ on the shape measurement and the dispersion $\sigma_\varepsilon$ on the intrinsic ellipticity of the sources. It is defined as 
\begin{equation}
    \omega_n = \frac{1}{\sigma^2_\varepsilon + \sigma^2_n} \, .
\label{omega}
\end{equation}
Finally, $\langle Z \rangle_n$ and $\langle Z^2 \rangle_n$ are weight factors taking into account the redshift distribution of the background galaxies. A more detailed discussion of the technique used to compute these quantities is deferred to Section \ref{analysis}. Here, we just mention that they are closely related to the critical surface mass density $\Sigma_\mathrm{cr}$.

\noindent The sums in Eq. (\ref{shear-field}) are extended to the number of background galaxies. \\ 

\noindent To recover the dimensionless surface mass density distribution $\kappa$ of the cluster we minimise the action \citep{lombardi3} $A$, extended over the field of view $\textit{U}$,
\begin{equation}
    A = \int_\textit{U} ||\vec{\nabla}\kappa - \vec{u}|| \, \mathrm{d}^2 x \, ,
\label{action}
\end{equation}
where $\vec{u}$ is a suitable combination of the derivatives of $\gamma$, 
\begin{equation}
    \vec{u} = \begin{pmatrix}
    \gamma_{1,1} + \gamma_{2,2} \\
    \gamma_{2,1} - \gamma_{1,2}
    \end{pmatrix} \, .
\label{u-vec}
\end{equation}
Here, the notation $\gamma_{i,j}$ indicates the derivative of the $i$-th component of $\gamma$ with respect to the $j$-th coordinate. Eq. (\ref{action}) could be inverted to immediately recover the convergence by solving the corresponding linear equation
\begin{equation}
    \vec{G} \kappa = \vec{u} \, ,
\label{sparse}
\end{equation}
where $\vec{G}$ is a sparse matrix implementing the operation of gradient.\\

\noindent Finally, since $\kappa$ is dimensionless, we obtain a dimensional surface mass distribution $\Sigma$ following this procedure: as stated more precisely in Section \ref{analysis}, $\kappa$ is estimated by assuming a fictitious source at redshift infinite, hence we recover $\Sigma$ by evaluating the critical surface mass density at the same redshift.

\section{Magellan observations}
\label{Data}

In this section, we present the data collected by the Subaru/Suprime-Cam and Magellan/MegaCam cameras we worked on, the identification of astronomical objects, and their morphological classification. We also discuss the selection adopted to identify cluster members and background galaxies. 

\subsection{Data description}

\begin{figure*}
    \centering
    \includegraphics[width=1.02\linewidth]{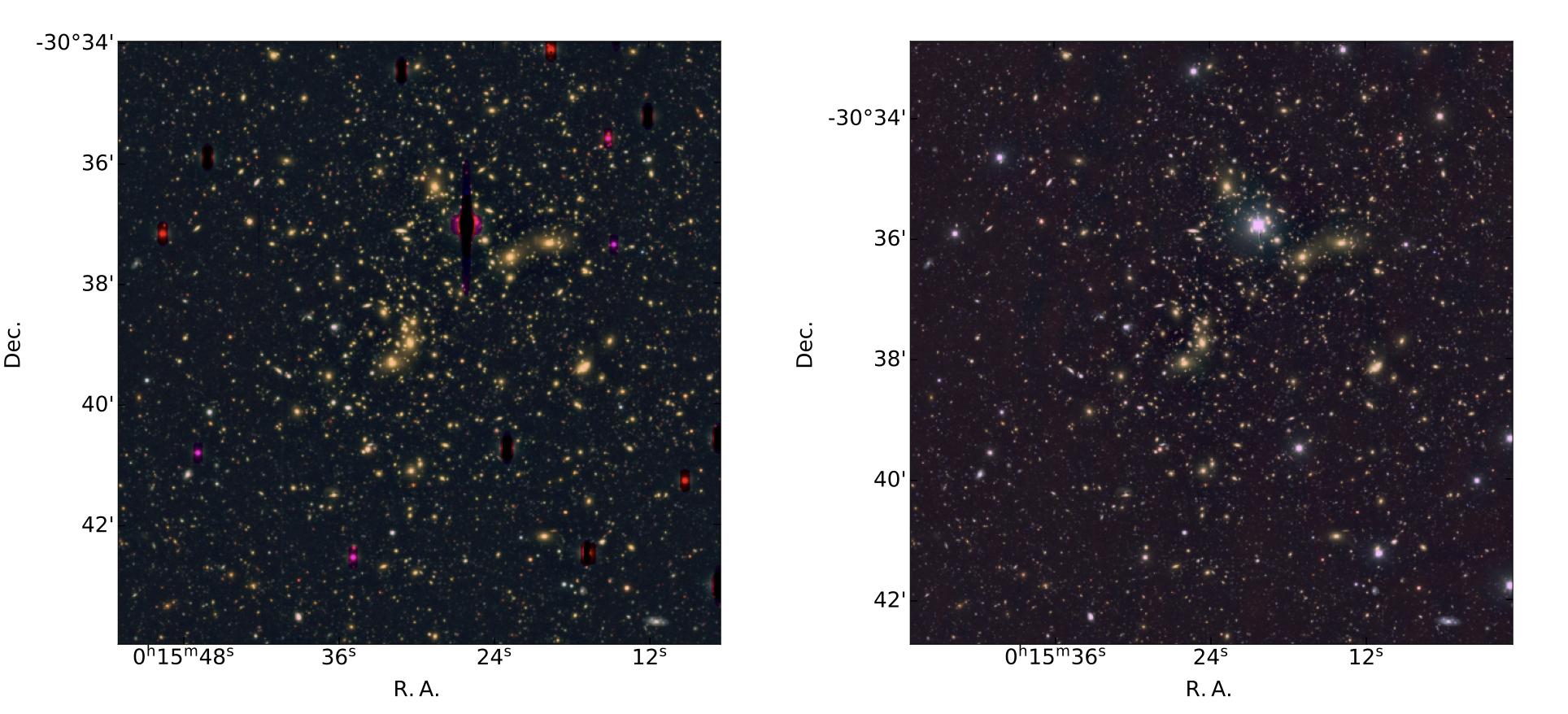}
    \caption{An extract of the field of view analysed centred on the galaxy cluster Abell 2744 as observed with the two facilities: the left panel depicts a color composite Subaru/Suprime-Cam $B$$R_c$$z'$ image, whereas the right panel is the color composite Magellan/MegaCam $g$$r$$i$ one.}
    \label{fig:abbello}
\end{figure*}

In our work, we applied two brand-new softwares for the PSF reconstruction and the measurement of the shapes of the background sources. For this reason, we first tested them on Subaru/Suprime-Cam \citep{subaru} $BR_cz'$ imaging data, collected on 2013 July 14-15, covering a field of view of approximately $24' \times 27'$. After this preliminary study, we used the pipeline on new data collected on 2018 September 7–8 \citep[see][]{treu} with the MegaCam camera \citep{mcleod}. This camera is located at the $f/5$ focus of the 6.5-m Magellan 2 Clay telescope at Las Campanas Observatory, Chile, and its focal plane is composed of $36 2048 \, \mathrm{px} \times 4608 \, \mathrm{px}$ CCDs which cover approximately a $25' \times 25'$ field of view, with a pixel scale of $0.08''$. The observations were carried out in the three filters $g$, $r$, and $i$. The data collected by both facilities were reduced following the procedure described in \citet{nonino} to create a co-added mosaic of images. The reduction steps include the subtraction of bias images, the application of flat-field corrections, the masking of bad and/or saturated pixels and artifacts. For further details, the reader is referred to \citet{nonino}. The co-added Magellan images were anchored to the Gaia-DR3 astrometric solution \citep[see][]{paris}. For each band, an effective $\simeq 31' \times 33'$ image was processed with the software $\mathrm{SWarp}$ \citep{bertin2002} in order to stack the single exposures on a common sky coordinate grid. For each band, the reduction pipeline also provided a weight map that we later used for our analysis. The observation details for both Subaru/Suprime-Cam and Magellan/MegaCam are listed in Table \ref{tab:filter}. In particular, we approximately estimated the PSF size by modelling each star with an isotropic Gaussian profile and by taking the median of the resulting best-fit full width at half maximum (FWHM) sizes for each band (see later). We used these estimates in order to produce a composite image, given by as a suitably weighted average of the single-band observations. Figure \ref{fig:abbello} is a RGB color composite image showing the central $10' \times 10'$ region of the field of view, obtained with both the Subaru/Suprime-Cam $BR_cz'$ data (left) and the Magellan/MegaCam $gri$ imaging (right). The results discussed and presented in the following refer to the analysis performed on the Magellan imaging, despite a comparison on the findings obtained with the two datasets is given in Section \ref{discussion}.

\begin{table}[!ht]
\caption{The specifics of the observations performed with the Subaru/Suprime-Cam and the Magellan/MegaCam cameras.}
\label{tab:filter}
\centering
\begin{tabular}{ccccc} 
 \hline \hline
  Instrument & Filter & Exp. time  & $m_\mathrm{lim}$\tablefootmark{a} & PSF size\tablefootmark{b}\\ 
&  & [$\mathrm{10^4 \, s}$] & ($\mathrm{AB \, mag}$) & [$\mathrm{''}$] \\
 \hline
& $g$ & 2.04 & 28.7 & 0.69 \\  
MegaCam & $r$ & 1.20 &  28.4 & 0.66 \\ 
 & $i$ & 4.80 &  27.8 & 0.63 \\ 
\hline
 & $B$ & 0.21 & 27.9 & 1.00 \\  
Suprime-Cam & $R_c$ & 0.31 &  27.7 & 1.04 \\ 
 & $z'$ & 0.36 &  27.5 & 0.79 \\ 
\hline
\end{tabular}
\tablefoot{
\tablefoottext{a}{$m_\mathrm{lim}$ denotes the $5 \sigma$ limiting magnitude evaluated within a Kron-like aperture.}
\tablefoottext{b}{The PSF size was estimated as the median of the detected stars' FWHM size distribution.}
}
\end{table}

\subsection{Source identification}

\begin{figure}
    \centering
    \includegraphics[width=1.02\linewidth]{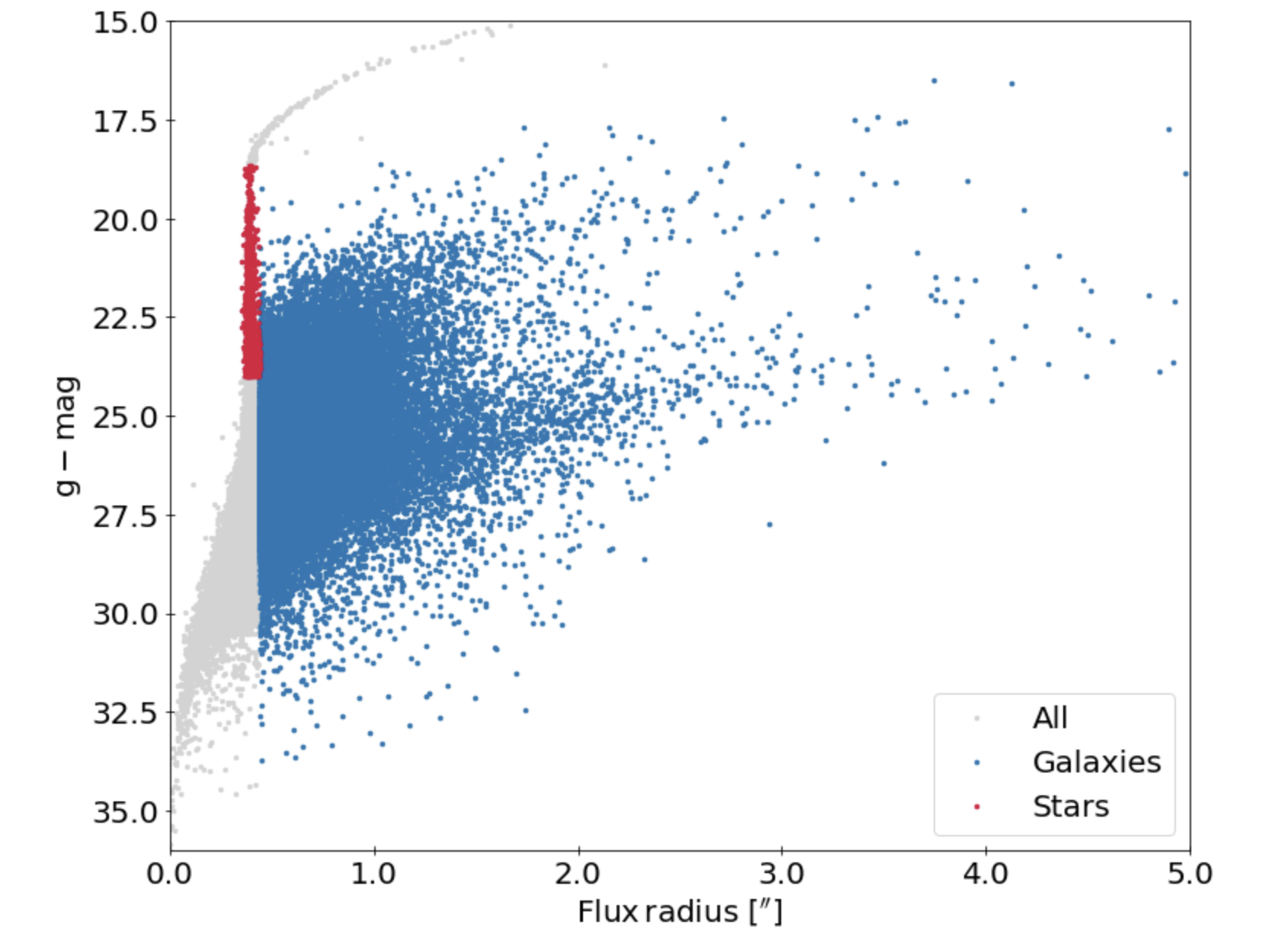}
    \caption{The classification of the sources between galaxies (blue) and stars (red) after the comparison of the three bands from the Magellan data, as illustrated in the text, in a magnitude vs. size diagram. Also shown are all the detected sources (in grey). The values of the magnitude and flux radius reported here refer to the $g$ band.}
    \label{fig:mag-rad}
\end{figure}

We first identified the astronomical sources located in the field of view with the software $\mathrm{SExtractor}$ \citep{sex} in the dual-image mode. The detections were carried out on the Magellan/MegaCam $gri$ composite image, because of their superior depth and resolution, whereas the measurements of the physical properties of the sources (e.g., size and magnitude) were performed on the single-band images individually. 117001 sources were thus recognised. We later classified them as galaxies and stars in the three bands independently, by studying their distribution in size and magnitude. As a reference for the size of the objects we used the \texttt{flux radius} measured by SExtractor, i.e., the radius of the isophote enclosing 40\% of the total luminosity of the source. In a magnitude vs. size diagram, stars occupy a well-defined vertical region of approximately constant radius, whereas galaxies are broadly distributed both in size and flux. After removing the sources associated with saturated pixels, we identified as galaxies those objects recognised as such in at least one of the three bands, and that were not classified either as stars or as saturated objects in the other filters. A similar procedure was followed to identify the stars. In our field of view, we thus identified 695 objects as stars and 86945 as galaxies. This corresponds to an average number density of $\simeq 84 \, \mathrm{galaxies/arcmin^{-2}}$. Figure \ref{fig:mag-rad} shows the distribution in size and magnitude of both stars and galaxies in the $g$ band; similar results were found for the other filters. It is worth mentioning the lack of the so-called brighter-fatter effect up to magnitude 19, as emerges from Figure \ref{fig:mag-rad}: the region occupied by the unsaturated stars is vertical, i.e., they have an approximately constant flux radius irrespective of their luminosity. Saturated stars have been explicitly ignored in our analysis and are shown as grey dots in Figure \ref{fig:mag-rad}.

\subsection{Identification of cluster and background galaxies}

\begin{figure*}
    \centering
    \includegraphics[width=1.02\linewidth]{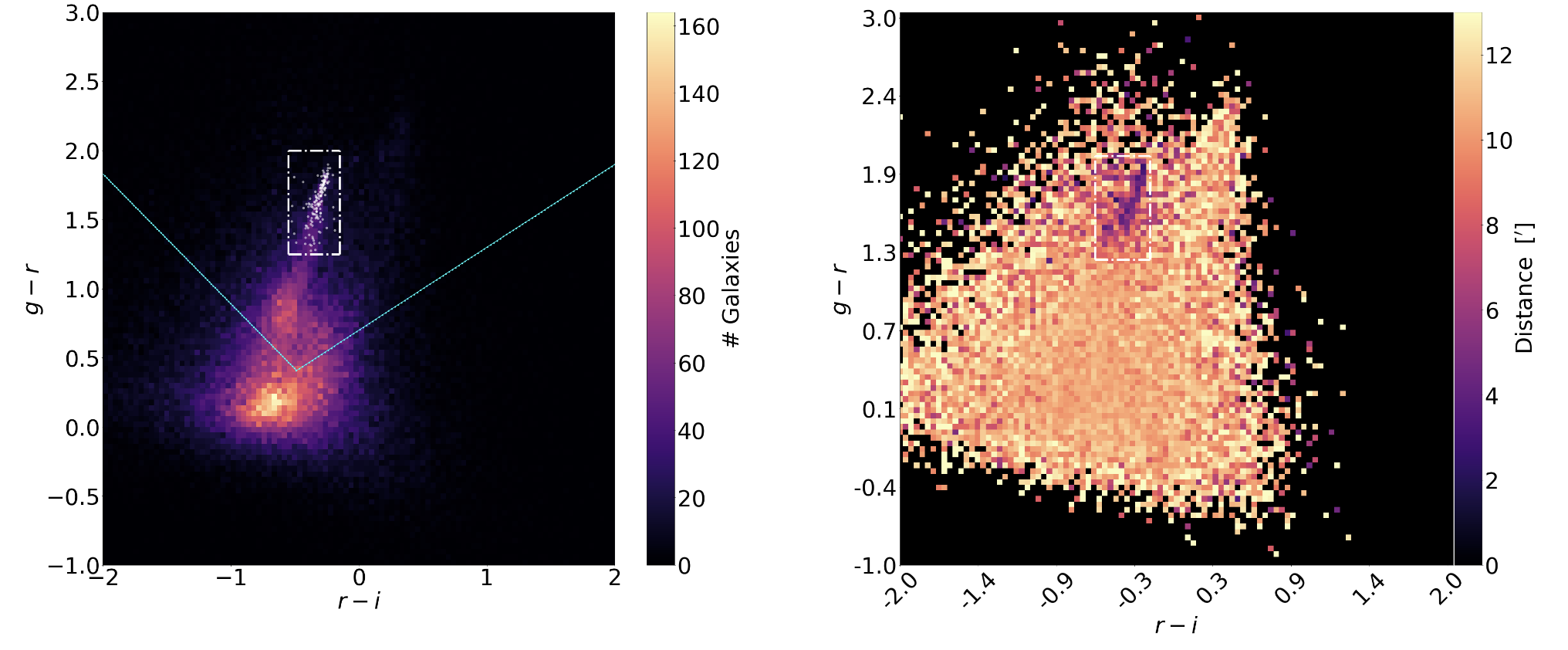}
    \caption{The left panel is the color-color (CC) diagram $g - r$ vs. $r - i$, whereas the right panel displays the distribution of the mean distances of the galaxies from the centre of the image in the same space. The rectangular box identified with the white lines in the upper-central corner of both panels identifies the region where the cluster was estimated to lie in the CC space. In the left panel, the white dots identify the 201 spectroscopically confirmed cluster members. The cyan lines in the left panel denote the region below which we assumed the background galaxies to lie.}
    \label{fig:col-col-dist}
\end{figure*}

\begin{figure}[h!]
    \centering
    \includegraphics[width=1.02\linewidth]{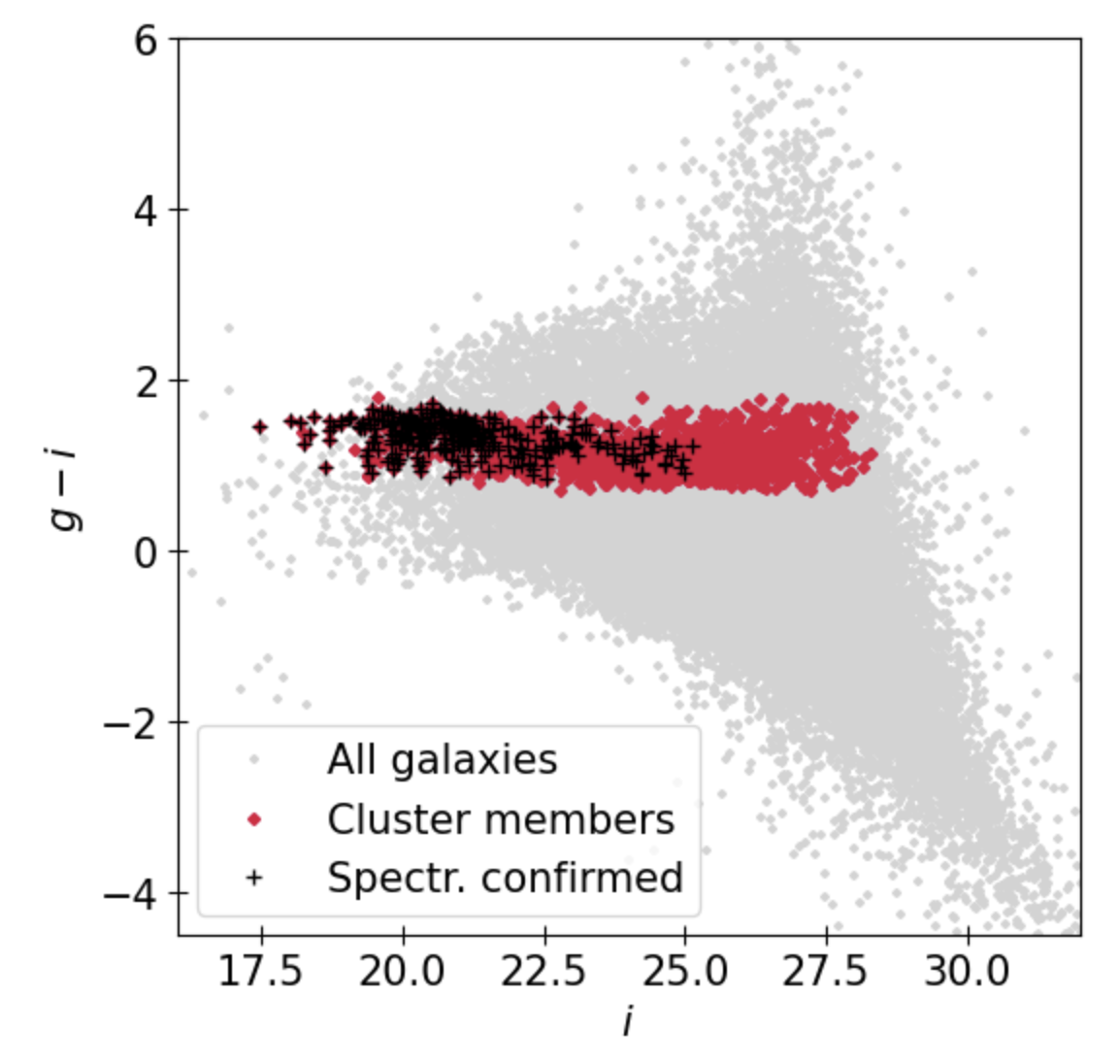}
    \caption{Color-magnitude diagram $g-i$ vs. $i$ for all the galaxies (grey) and for those we identified as cluster members (red). The latter define a clearly visible red cluster sequence. The black crosses identify the 201 spectroscopically confirmed cluster members.}
    \label{fig:rcs}
\end{figure}

Afterwards, we distinguished the galaxies between cluster members and background ones by adopting the procedure outlined by \citet{mede2010, mede2018} and successfully applied in several WL analysis of galaxy clusters \citep[see, e.g.,][]{jauzac2012, mede2013, mede2016}. We produced a color-color (CC) $g - r$ vs. $r - i$ diagram (the left panel of Figure \ref{fig:col-col-dist}) and evaluated the mean distance of all the galaxies from the cluster centre in a given CC cell (see the right plot of Figure \ref{fig:col-col-dist}). We considered as the centre of the system the brightest cluster galaxy (BCG). The upper-central region (identified by the white contours in both panels) is mainly populated by galaxies having the smallest mean distance from the centre of the cluster. It can be seen that this region corresponds to a local over-density in the CC space. We therefore selected the galaxies residing in this region as potential cluster members, and further restricted to those lying within $1.2 \, \mathrm{Mpc}$ (i.e., $6'$ at the redshift of the cluster) from the centre of the image. We identified in this way 1477 galaxies likely belonging to the cluster. These galaxies occupy a well defined red cluster sequence in the color-magnitude diagram depicted in Figure \ref{fig:rcs}. \\

We verified the purity and completeness of our cluster sample by comparing it to the spectroscopic redshift catalogue used by \citet{bergamini2023a}, who performed a SL study of the same galaxy cluster based on JWST/NIRCam imaging and Very Large Telescope (VLT) Multi Unit Spectroscopic Explorer (MUSE) spectroscopy. The catalogue contains the spectroscopic redshifts of 2397 objects, from \citet{braglia}, \citet{owers} and \citet{richard}. To make a fair comparison with our sample, we considered only the galaxies lying in the redshift range $(0.28, 0.34)$ (which corresponds to a rest-frame velocity of the galaxies in the cluster of $\pm 6000 \, \mathrm{km/s}$ around the median redshift of the cluster) and within $6'$ from the cluster centre. Out of the 710 galaxies present in both datasets, 201 of them were classified by the authors as cluster members and satisfied the distance criterion described above, and we correctly identified 150 of them. We quantified the goodness of our classification by estimating the purity and the completeness, defined as
\begin{equation}
    \mathrm{purity} = \frac{TP}{TP + FP} \, ,
\end{equation}
 and 
\begin{equation}
    \mathrm{completeness} = \frac{TP}{TP + FN} \, .
\end{equation}
Here, $TP$, $FP$, and $FN$ denote the true positives, the false positives and the false negatives, respectively. We achieved a purity of \textbf{$\simeq 84\%$} and a completeness of \textbf{$\simeq 87\%$}. The spectroscopically-confirmed cluster members are depicted as white dots in the left panel of Figure \ref{fig:col-col-dist}.\\

\begin{figure}[h!]
    \centering
    \includegraphics[width=1.02\linewidth]{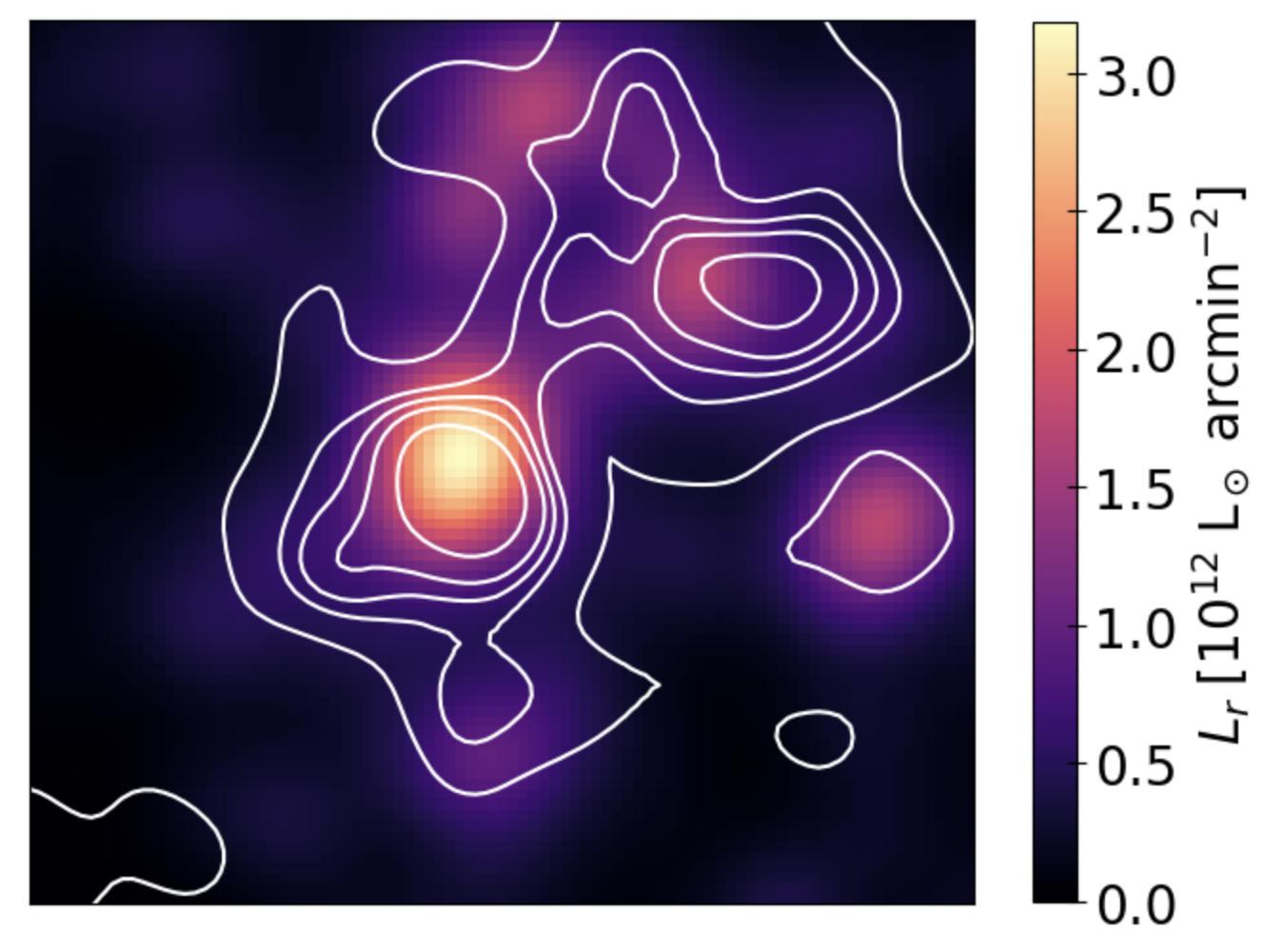}
    \caption{The $7' \times 7'$ extract of the surface brightness distribution of the cluster in the $K$-corrected $r$ band, with overlaid (in white), the contours of the total surface mass distribution at 2.5, 5, 6, 7 and 9 $\sigma_k$ (see Section \ref{analysis}).}
    \label{fig:lum-map}
\end{figure}

With the 1477 galaxies selected earlier to be potential cluster members we reconstructed the surface brightness distribution of the cluster. We took into account the $K$-correction \citep[see, e.g.,][]{hogg} by following the procedure in \citet{beare}, that allowed us to express the $K$-corrected magnitudes as a suitable combination of the available colors. The luminosity distribution in the $K$-corrected $r$ band is depicted in Figure \ref{fig:lum-map}, where the contour lines of the recovered surface mass density (see later) are also overlaid. \\

Finally, we isolated the foreground galaxies from the background ones by identifying the region corresponding to an absolute maximum in the galaxy number density in the centre of the CC diagram (left panel of Figure \ref{fig:col-col-dist}), which is mainly populated by foreground sources: we thus obtained a catalogue consisting of 54614 potentially background galaxies. The cyan lines in the left panel of Figure \ref{fig:col-col-dist} denote the region below which we assumed the background galaxies to lie.  

\section{WL analysis}
\label{analysis}

\subsection{PSF reconstruction}

Before measuring the ellipticities of the background galaxies, we reconstructed the spatial variation of the point spread function (PSF) all over the field of view. Indeed, the PSF introduces an artificial shape distortion, which has to be accurately measured and properly corrected for. For this purpose, we used $\mathrm{mccd}$ \citep{mccd}, a module of the pipeline shapepipe \citep{guinot, ferrens} we applied for our WL analysis. Differently from other softwares, mccd is capable of handling multiple CCDs simultaneously, thus handling discontinuities at the CCDs boundaries and providing a single continuous PSF field, instead of producing a PSF model for each CCD separately. The point spread function is described by the software in terms of a model, specified by the user, splitted into a global and a local term. The latter describes the PSF variation over a single CCD as a function of the pixels coordinates, whereas the former takes into account all the CCDs. mccd operates on stamps of the stars to measure their shapes and thus determine the parameters of the model, which are later interpolated at the coordinates of the galaxies. In our work, both the global and the local terms coincided, since, in order to simplify the weak lensing pipeline, we directly operated on the single-band co-added images, instead of working on individual frames. This may result in a limitation in the use of the code; however, we opted for a less complex methodology, and given the results we obtained (see later), we are satisfied with the results obtained with mccd. We applied the software on the three single-images separately, hence obtaining three different PSF maps. \\

\begin{figure*}
    \centering
    \includegraphics[width=1.0\linewidth]{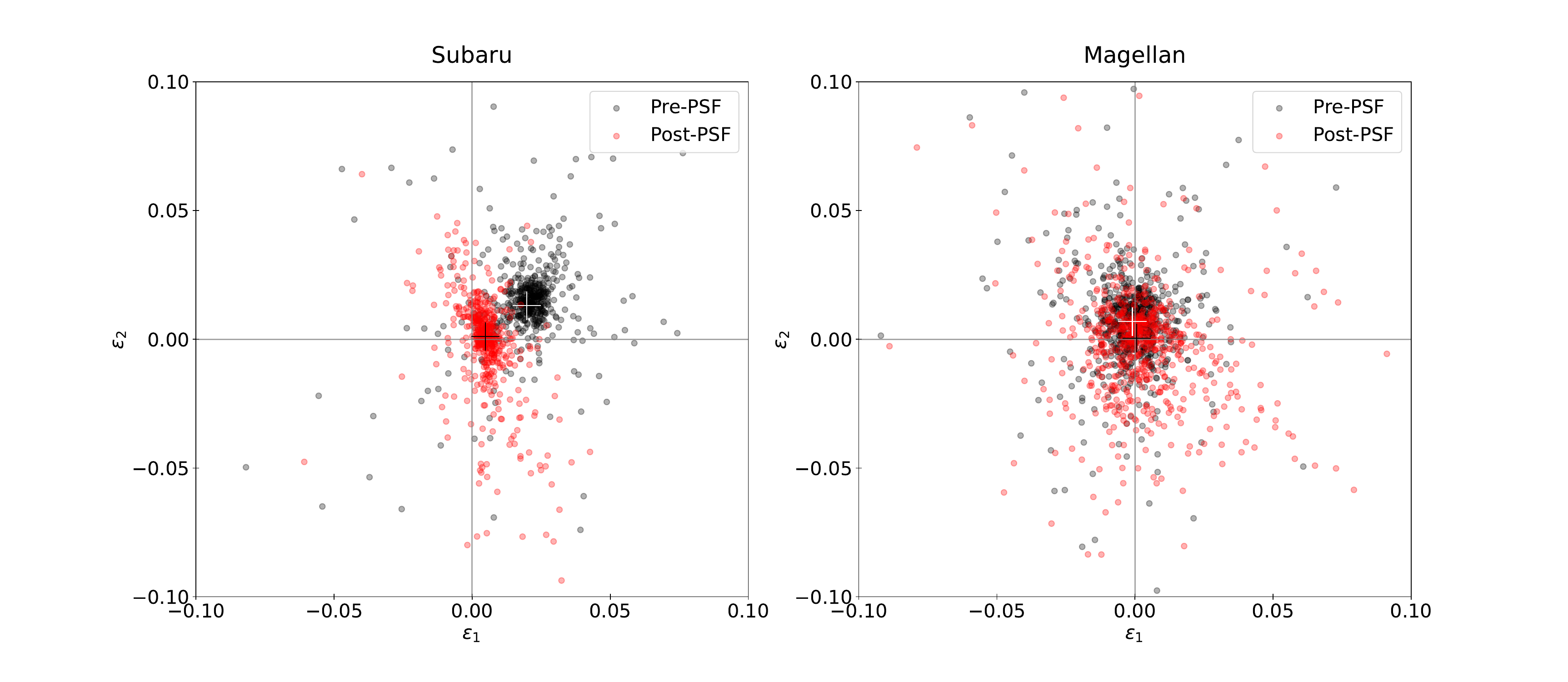}
    \caption{Ellipticity distribution of the stars on the field of view before (in black) and after (in red) the PSF correction. The white (black) cross represents the mean of the distribution before (after) the correction. The left (right) panel shows the results for the Subaru (Magellan, respectively) facility.}
    \label{fig:mccd}
\end{figure*}

We quantified the goodness of our PSF reconstruction by measuring the difference between the ellipticity $\varepsilon = \varepsilon_1 + i \varepsilon_2$ of the observed stars and the one obtained after correcting for the PSF reconstruction at the same locations. The result is depicted in Figure \ref{fig:mccd}, that shows the ellipticity distribution both before (black) and after (red) the PSF correction, as well as the position of the means of the distributions (in white and black, respectively), obtained with the Python-based huber estimator \citep{huber}. The results obtained with both the Subaru/Suprime-Cam (left panel) and the Magellan/MegaCam (right panel) cameras are shown. As far as the Magellan results are concerned, before correcting for the PSF, we estimated mean values of $\langle \varepsilon_1 \rangle_{before} = (-0.95 \pm 9.28) \times 10^{-3}$ and $\langle \varepsilon_2 \rangle_{before} = (0.69 \pm 1.37) \times 10^{-2}$. After the correction, we recovered mean values of $\langle \varepsilon_1 \rangle_{after} = (0.59 \pm 1.06) \times 10^{-3}$ and $\langle \varepsilon_2 \rangle_{after} = (0.46 \pm 1.53) \times 10^{-3}$. 

\subsection{Shear measurement}

We thus proceeded with the measurements of the shapes of the background galaxies with the module ngmix \citep{sheldon1, sheldon2} of the aforementioned shapepipe pipeline, which follows a model-fitting approach. For each galaxy, a stamp centred on it must be supplied. We used a square box with a size of six times the highest value of the flux radius (among the three bands) corresponding to that galaxy. The algorithm first fits the PSF associated with it $-$ that has been reconstructed in the previous passage $-$ as a mixture of $m$ co-axial Gaussian distributions, with the integer $m$ specified by the user. Afterwards, the galaxy shape is measured: galaxies are initially described in terms of a parametrised surface brightness distribution; later, the resulting image is sheared and the previously fitted PSF is applied. Then, ngmix applies a least-squares method based on the Levenberg-Marquardt algorithm to fit the original image with the model and returns the best-fit model parameters, including the two components of the shear. This operation is run once, then the process is bootstrapped several times, such that an entire strip of best-fit values is returned, as well as their standard deviations and covariance matrix. Throughout the analysis, we set $m = 2$ and adopted a simple exponential surface brightness profile to model the galaxies, since we were interested in their shapes only. Before opting for this choice, we had run ngmix both on simulated galaxies and on Subaru/Suprime-Cam data by using different models. Extensive tests on different combinations of galaxy shapes have shown that, for galaxies with angular sizes not much larger than the PSF size, as in our situation, a fit consisting of a single exponential profile is much more robust and reliable than a fit obtained by using a combination of exponential and de Vaucoleurs profiles. This is likely associated with the presence of degeneracies when a limited amount of information is available. Therefore, we opted for a pure exponential profile.

This algorithm does not measure the galaxy shapes correctly if the image provided to the software contains two or more objects, since it is not capable of disentangling multiple profiles. Therefore, we initially removed from our background sample the galaxies with overlapping isophotes, by exploiting the segmentation map produced by SExtractor. Furthermore, we removed from the subsequent analysis the galaxies lying close to bad and/or saturated pixels, stars of our Galaxy and the edges of the field of view. We therefore applied ngmix to measure the shape of the 20519 remaining background galaxies. To do so, we adopted the so called multi-band configuration, i.e., the measurements were carried out on an image created by stacking three stamps of each galaxy in the three filters individually. For each galaxy, the stamp was obtained on the three mosaic co-added images corresponding to the three different filters. We also supplied each single-band stamp a corresponding box in the weight map and in the PSF map. In case of galaxies with a signal-to-noise ratio (S/N) measured by SExtractor less than 10 in a given band, the weight box in that filter was set to zero when creating the multi-band stamp, as to force the algorithm not to use that filter. \\

After these measurements, we further cleaned our sample before reconstructing the shear field all over the field of view. We removed the galaxies whose shear components were both measured with a standard deviation higher than 0.7, those whose centre (as measured with SExtractor) differed from the one fitted with ngmix by more than $0.75''$ and those with a S/N evaluated by the algorithm less than 10. This left us with $N_b = 13942$ background galaxies to compute the total mass distribution with, corresponding to $\simeq 15 \, \mathrm{galaxies/arcmin^2}$. We thus reconstructed the shear field $\gamma(\vec{x})$ by applying Eq. (\ref{shear-field}). To do so, we pixelised the field of view onto a $480 \, \mathrm{px} \times 500 \, \mathrm{px}$ grid (corresponding to a pixel scale of $0.1'$) and introduced an isotropic two-dimensional Gaussian weight function $K$ with standard deviation equal to $0.5'$. We evaluated the dispersion on the intrinsic ellipticity appearing in Eq. (\ref{omega}) by using our shape measurements, as 
\begin{equation}
    \sigma^2_\varepsilon = \frac{\sum_n^{N_b} (\varepsilon^2_n + \sigma^2_n)}{2 \, N_b } \, ,
\end{equation}
where, for each galaxy, $\sigma^2_n$ is given by the sum in quadrature of the uncertainties of the two components of the ellipticity returned by ngmix. We obtained $\sigma_\varepsilon = 0.25$. The weights $\{\langle Z \rangle_n\}$ and $\{\langle Z^2 \rangle_n\}$ appearing in Eq. (\ref{shear-field}), as suggested previously, depend on the redshift distribution of the background galaxies. Since we did not have at our disposal this information, we proceeded in the following way to evaluate them. We derived the redshift distribution from the photometric redshift catalogue centred in the Hubble Deep Field - North by \citet{yang}. We first selected the sources in the catalogue being recognised as galaxies, and then applied the same selection criteria described in the previous section to determine the background ones. If $z_j$ denotes the redshift of the $j$-th catalogue galaxy with magnitude $m$ in the interval $[m_i, m_i + 0.5)$ in a given band and $N_i$ the number of catalogue galaxies in the same bin, we evaluated the $\langle Z \rangle_i$ and the $\langle Z^2 \rangle_i$ corresponding to the $i$-th bin by using the equation,
\begin{equation}
    \langle Z \rangle_i = \left \langle \frac{\Sigma_\mathrm{cr, \infty}}{\Sigma_\mathrm{cr}}\right \rangle _i = \frac{1}{N_i} \sum_{j=1}^{N_i} \frac{\Sigma_\mathrm{cr, \infty}}{\Sigma_\mathrm{cr} (z_j)} \, ,
\end{equation}
and
\begin{equation}
    \langle Z^2 \rangle_i = \left \langle \frac{\Sigma^2_\mathrm{cr, \infty}}{\Sigma^2_\mathrm{cr}}\right \rangle _i = \frac{1}{N_i} \sum_{j=1}^{N_i} \frac{\Sigma^2_\mathrm{cr, \infty}}{\Sigma^2_\mathrm{cr} (z_j)} \, .
\end{equation}
In the previous equations, $\Sigma_\mathrm{cr, \infty}$ is the critical surface mass distribution given in Eq. (\ref{crit-sigma}) for a fictitious source at redshift infinite. We thus assigned each of our background galaxies having a magnitude in the $i$-th bin the corresponding $\langle Z \rangle_i$ and $\langle Z^2 \rangle_i$ values. As a reference band, we opted for the $r$ filter. Typical values of the variance of $\langle Z \rangle_i$ and $\langle Z^2 \rangle_i$ are around unity for all magnitude bins.

\subsection{Cluster mass reconstruction}

We thus reconstructed the convergence $\kappa$ and then the total surface mass distribution $\Sigma$ through Eq. (\ref{action}), by applying Eq. (\ref{sparse}). We first constructed $G$ and then solved the linear equation using a least-squares method to obtain $\kappa$ and hence $\Sigma$. {We applied the iterative procedure described in \citet{seitz}, which reached convergence in five steps. We also estimated a statistical error $\sigma_k$ on the reconstruction of $\Sigma$ through a reshuffling procedure \citep{lombardi1}, i.e., we produced further 120 mass density maps, each time shuffling the coordinates of the background galaxies, and then evaluated the rms error $\sigma_k$ on the resulting 120 realisations of $\Sigma$. Figure \ref{fig:sigma-picchi} depicts the contour levels of $\Sigma$ over the central $6' \times 6'$ region of the $g$ band image.\\ 

\begin{figure}
    \centering
    \includegraphics[width=1.0\linewidth]{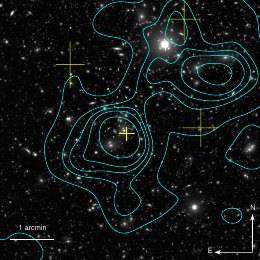}
    \caption{The central $6' \times 6'$ region of the $g$-band image with overlaid the contour levels of the total surface mass distribution at 2.5, 5, 6, 7 and 9 $\sigma_k$. The yellow crosses denote the density peaks identified by \citet{mede2016}.}
    \label{fig:sigma-picchi}
\end{figure}

We further looked for potential systematic effects by evaluating the so-called $B$-modes. The two components of the shear can indeed be decomposed into two parts, the $E-$ (curl$-$free) and the $B-$ (divergence$-$free) modes. In the context of gravitational lensing, the $E-$mode signal is related to the surface mass distribution, whereas the $B-$mode signal is identically zero \citep{umetsu2020}. The presence of non-null $B-$modes can thus be used as a test for the presence of systematic effects. To do so, we evaluated the lensing signal $\Sigma$ after rotating each component of the shear of an angle equal to $\pi/4$, i.e., mapping $\gamma_1 \rightarrow -\gamma_2$ and $\gamma_2 \rightarrow \gamma_1$. We followed the same procedure outlined before and then estimated the rms error on the distribution found in this way by means of a similar reshuffling procedure. The resulting $B-$mode thus obtained is consistent with zero.\\

Finally, as in all WL analyses, our mass reconstruction is affected by the so-called mass-sheet degeneracy, i.e., $\Sigma$ can be determined up to an additive constant $c$ \citep[for a review, see again][]{bartelmann}. Indeed, as can be seen in Eq. (\ref{action}), an additive constant does not modify the gradient, and therefore the minimisation of the action. To deal with this degeneracy and obtain an estimate of $c$, we modeled the main density peak of our surface mass distribution (the one in the south-east part of the cluster, see later) in terms of a simple softened isothermal ellipsoid \citep[SIE; see, e.g.,][]{keeton}, depending, among the other parameters, on the additive constant $c$. 

To determine the free parameters of the model, $\lambda$, we performed a Bayesian inference. We expressed the likelihood as the product of bi-dimensional Gaussian distributions $\mathcal{N}(\Sigma(\vec{x}) - \Sigma_0(\vec{x}|\lambda), \sigma^2_k)$, where $\Sigma(\vec{x})$ and $\Sigma_0(\vec{x}|\lambda)$ are the reconstructed and modelled total surface mass distributions given the parameters $\lambda$ at location $\vec{x}$, respectively, and $\sigma^2_k$ is the value of the variance map at the same coordinate. We only restricted to the annulus with internal radius equal to $\sim 3 \, \mathrm{Mpc}$ and external radius equal to $\sim 4 \, \mathrm{Mpc}$, where the radial density profile of the cluster is supposed to vanish. We inferred the parameters assuming as a prior a uniform distribution over a suitable subset in the parameters’ space. In particular, we left $c$ free to vary in the interval $[-1, 1] \times 10^{15} \, \mathrm{M}_\odot/\mathrm{Mpc}^2$. We sampled the posterior distribution with the nested sampling Monte Carlo algorithm Ultranest \citep{buchner}. We found $c = (4 \pm 1) \times 10^{12} \, \mathrm{M}_\odot/\mathrm{Mpc}^2$. Figure \ref{fig:comp} shows the cumulative mass profiles, obtained by averaging the surface mass distribution on circular apertures. We adopted as the centre of the mass profile the centre of the cluster indicated at the end of Section \ref{intro}. The plot also depicts the results from the preliminary WL analysis carried out on Subaru/Suprime-Cam $B$$R_c$$z'$ data, on which we first applied the pipeline described above. 

\begin{figure*}
\begin{minipage}{0.7\textwidth}
  \centering
   \includegraphics[width=1.02\linewidth]{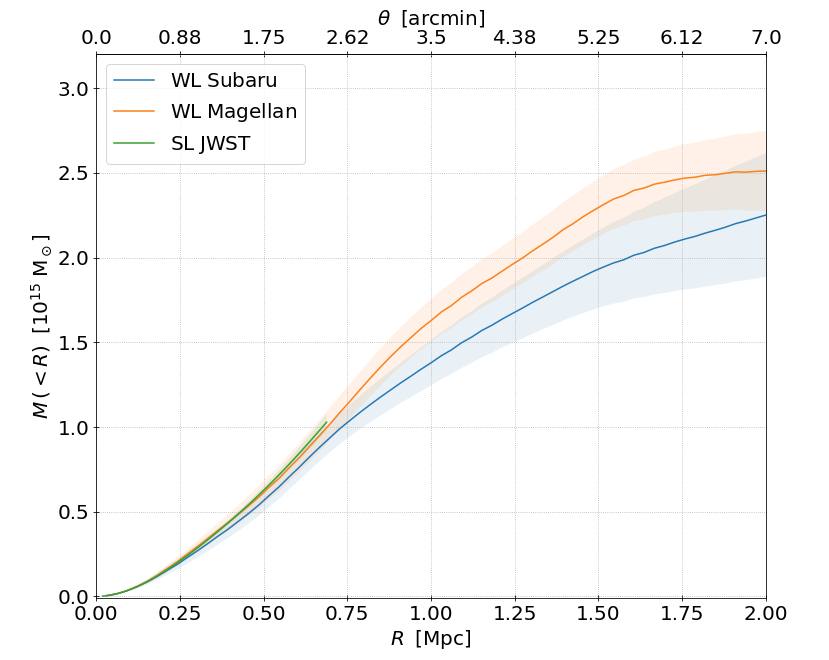}
\end{minipage}
\begin{minipage}{0.3\textwidth}
    \caption{Cumulative radial total mass profile of the cluster as obtained with Subaru/Suprime-Cam (blue) and Magellan/Megacam (orange) imaging. Also shown is the cumulative profile (green) in the core of the cluster as a result of the SL analysis based on JWST \citep{bergamini2023b}. The shaded area denotes the error band corresponding to the 68\% confidence level.}
    \label{fig:comp}
\end{minipage}
\end{figure*}

\section{Discussion and results}
\label{discussion}

\subsection{Comparison between SL \& WL}

We presented in the previous sections the WL analysis we performed to reconstruct the total mass distribution of the galaxy cluster Abell 2744. At a distance of $2.35 \, \mathrm{Mpc}$ from the south-east BCG, we recovered a total mass of $(2.56 \pm 0.26) \times 10^{15} \, \mathrm{M}_\odot$, which is slightly larger the previous WL study of the same cluster by \citet[][]{mede2016}, who found a value of the total mass within the same aperture of $(2.06 \pm 0.42) \times 10^{15} \, \mathrm{M}_\odot$, despite being consistent within one rms. Moreover, the total mass enclosed within $1 \, \mathrm{Mpc}$, equal to $(1.68 \pm 0.13) \times 10^{15} \, \mathrm{M}_\odot$, is in very good agreement with the estimate obtained by \citet{boschin}, equal to $(1.4-2.4) \times 10^{15} \, \mathrm{M}_\odot$ within the same radius, who performed a dynamical analysis of the cluster based on New Technology Telescope (NTT) European Southern Observatory (ESO) Multimode Instrument \citep[EMMI,][]{emmi} spectrography. Additionally, our result is consistent with the finding of the afore-mentioned combined SL\&WL analysis by \citet{jau}: they found a total mass enclosed within $1.3 \, \mathrm{Mpc}$ from the BCG equal to $(2.3 \pm 0.1) \times 10^{15} \, \mathrm{M}_\odot$, that is perfectly consistent with the total mass enclosed within the same radius we found, $(2.1 \pm 0.2) \times 10^{15} \, \mathrm{M}_\odot$. To further corroborate the validity of our results, we compared our findings with the SL study by \citet{bergamini2023b} based on new deep, high-resolution JWST imaging, after extrapolating their best-fit model up to $700 \, \mathrm{kpc}$. To make a fair comparison between the outcomes from the two different probes, we downgraded the best-fit SL surface mass distribution to the same resolution of our WL map. We then estimated the uncertainty on it by evaluating further 120 maps using different combinations of randomly extracted parameters from the SL model MCMC chains and evaluating the rms error. As can be seen in Figure \ref{fig:sigma-SL-WL}, where both the best-fit SL model and the WL surface mass distributions are overlaid on a JWST $rgb$ composite image, a good agreement emerges, as far as the southern and the right north-west density peaks are concerned. Only a slight discrepancy in the left north-west peak is present, likely due to the presence of a saturated star in the Magellan images (which is also visible in the JWST composite image), which we had to suitably mask out, thus losing shape information about the galaxies located in its neighborhood. We note that our WL study is capable of extrapolating the SL map out to large radii (up to few Mpc). As a further test, we evaluated the radial cumulative mass distribution from the SL probe $-$ by following the same procedure outlined above for the WL analysis $-$ which is displayed in Figure \ref{fig:comp}. Our findings are in a clearly excellent agreement within one $\sigma_k$.\\

\begin{figure}
    \centering
    \includegraphics[width=1.0\linewidth]{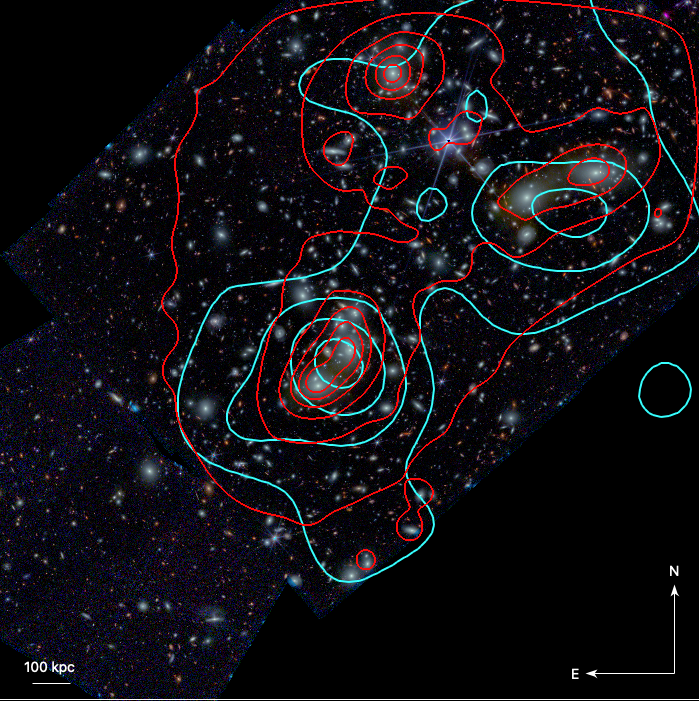}
    \caption{The central $900 \, \mathrm{kpc} \times 900 \, \mathrm{kpc} $ region of the composite JWST $rgb$ image with overlaid the contour levels of the surface mass distribution emerging from the SL \citep[][red]{bergamini2023b} and our WL (cyan) studies. The contour levels depicted are linearly spaced between $0.5 \times 10^{15} \, \mathrm{M}_\odot / \mathrm{Mpc^2}$ and $2.4 \times 10^{15} \, \mathrm{M}_\odot / \mathrm{Mpc^2}$}.
    \label{fig:sigma-SL-WL}
\end{figure}

Furthermore, our analysis reveals the presence of three density peaks having a S/N greater than 7: one, with higher density and a S/N ratio of 14.0, lies in the south-east part of the cluster, very close to the BCG, whereas the others are found in the north-west corner. All of them lie close to those observed in the SL surface mass distribution by \citet{bergamini2023b}. A slight overlap with the locations of the substructures detected in the  WL study by \citet{mede2016} (the yellow crosses in Figure \ref{fig:sigma-picchi}) is also observed. In particular, the south-east substructre detected in \citet{mede2016} lie well within our density peak, as well as the north-west one. This result suggests that the cluster is not relaxed, but is undergoing a complex merging process, as supported by the previously mentioned papers.\\ 

\subsection{Comparison between mass and luminosity distributions}

We compared the recovered surface mass distribution with the luminosity density of the cluster, as represented in Figure \ref{fig:lum-map}. As can be seen, there is a nice agreement between mass and luminosity, except for a slight misalignment as far as the north-west density peak is concerned. We explain this discrepancy as a consequence of the afore-mentioned effect due to the presence of the star we had to mask out. Furthermore, we estimated a mass-to-light ratio $M/L$ in the $K$-corrected $r$ band as the ratio between the total mass and the total luminosity $L_r$ within 200 kpc: we found $M(< 250 \, \mathrm{kpc})/L_r(< 250 \, \mathrm{kpc}) = 573 \pm 69 \, \mathrm{M_\odot/L_{\odot,r}}$, which is perfectly consistent with the results found by \citet{mede2016} with Subaru imaging. \\

\subsection{Comparison between Subaru and Magellan}

In the previous sections, we presented the results obtained with Magellan/MegaCam, after we tested both mccd and ngmix on the Subaru dataset, despite the latter being characterised by a worse PSF and depth. These features therefore impacted the quality of the analysis. Firstly, with Subaru/Suprime-Cam, we detected less galaxies, as can be seen in Figure \ref{fig:number-counts}, which displays the number counts per unit area referred to two filters, $R_c$ and $r$ for Subaru and Magellan, respectively. In particular, after applying the same CC method described previously, the average number density of background sources was slightly less than half of the one with Magellan ($\simeq 33 \, \mathrm{galaxies/arcmin^2}$). Moreover, after removing the galaxies unsuitable for the WL analysis, this density lowered to $\simeq 9 \, \mathrm{galaxies/arcmin^2}$. Secondly, we reported a significant difference as far as the shape measurements are concerned. Indeed, for the Subaru dataset only, we found a significant fraction of background galaxies with an ellipticity as measured by ngmix higher than 0.97 and an error on both the components of $\varepsilon$ higher than the unity, thus indicating the failure in fitting the galaxy shape due to the low S/N ratio. Additionally, ngmix fails in measuring, with a sufficiently low error, the shape of faint galaxies, as emerges from Figure \ref{fig:erroreSubMag}. The median error on $\varepsilon$ indeed increases with the apparent magnitude of the galaxies, with the errors on the Subaru imaging being steadily higher than those with Magellan imaging data. We therefore had to remove these sources from the subsequent analysis, which thus led to a result slightly different from the one obtained with Magellan. As a matter of fact, at a distance of $2 \, \mathrm{Mpc}$ from the BCG, the total mass is slightly lower, $(2.41 \pm 0.47) \times 10^{15} \, \mathrm{M}_\odot$, and characterised by a rms almost twice the one obtained with Magellan. Nevertheless, the two results agree with each other within one rms, as emerges from Figure \ref{fig:comp}.

\begin{figure}
    \centering
    \includegraphics[width=1.0\linewidth]{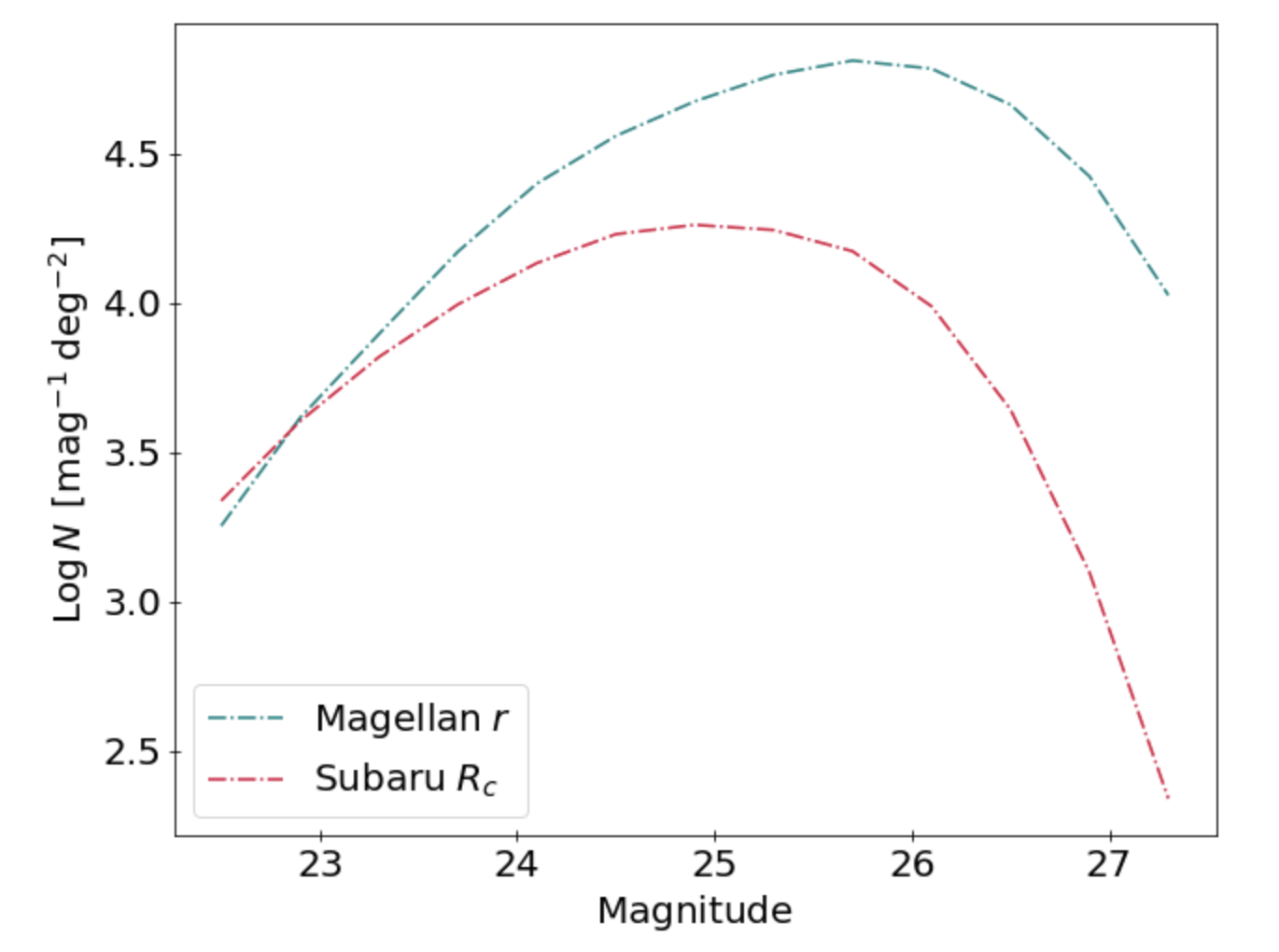}
    \caption{Number counts of the galaxies per unit area and magnitude for the Subaru (red) $R_c$ band and the Magellan (blue) $r$ filter.}
    \label{fig:number-counts}
\end{figure}

\begin{figure}
    \centering
    \includegraphics[width=1.0\linewidth]{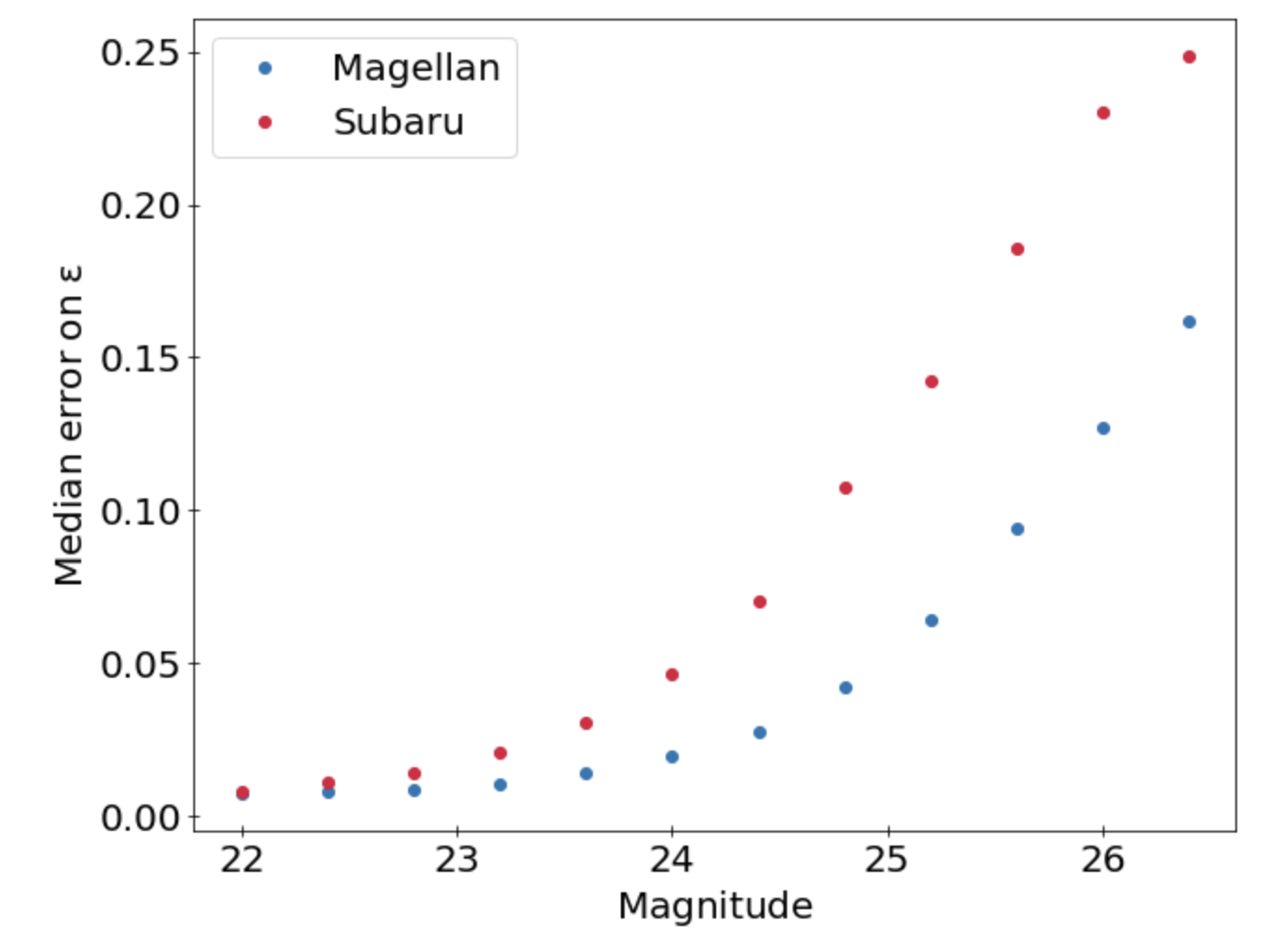}
    \caption{Distribution of the median error on the ellipticity as a function of the apparent magnitude of the background galaxies, for both the Magellan (blue) and the Subaru (red) datasets.}
    \label{fig:erroreSubMag}
\end{figure}

\subsection{Data contamination}

The results we obtained depend on several factors. Firstly, we classified the galaxies between cluster members, foreground and background ones by studying their distribution in a CC space only. An erroneous classification could lead to a dilution in the WL signal, and therefore influence the outcome of our analysis. We took care of this systematic by verifying the correctness of our classification through a comparison with the spectroscopic redshift catalogue by \citet{bergamini2023a}. Out of the galaxies present in both catalogues, we correctly identified the background sources with a purity of $\simeq 70\%$. It must be underlined, however, that the VLT/MUSE catalogue was obtained by analysing an area of few square arcmins, whereas our field of view covered an area of approximately $0.3 \, \mathrm{deg}^2$. To overcome this bias, it would be useful for future WL works the use of multi-band imaging data obtained from several bands in order to estimate photometric redshifts, thus allowing for a complementary method for the classification. These redshifts would also allow one to estimate the critical surface mass density in Eq. (\ref{crit-sigma}) for each background galaxy. This would lead to a more robust mass reconstruction. 

Similarly, we propose the use of other techniques, including the application of Convolutional Neural Networks (CNNs) \citep{lecun1989, lecun1998}. These algorithms are composed of several convolutional layers that are able to recognise complex features within the images on which they are applied. Recent works show how the CNNs are capable of achieving a high ($> 90 \%$) purity-completeness rate when tested on multiband imaging data \citep[see][]{angora}, thus making their application in upcoming WL studies promising. \\

\section{Conclusions}
\label{conclusions}

We have presented a WL analysis of the galaxy cluster Abell 2744 using new deep multi-band $gri$ imaging from Magellan/Megacam. For our study, we applied a pipeline based on two brand-new softwares, mccd and ngmix for  the PSF reconstruction and shape measurement, respectively. Differently from previous algorithms developed for this scope, both mccd and ngmix allow one to exploit the information from all the bands simultaneously. Moreover, mccd is capable of modelling the spatial variation of the point spread function of several CCDs, thus avoiding potential discontinuities at the boundaries. To test the robustness of these algorithms, we first applied them on multi-band $B$$R_c$$z'$ Subaru/Suprime-Cam imaging of the same cluster, covering an effective field of view of $\simeq 24' \times 27'$, and with a depth of $\simeq 27.9$ in the $B$ band. Afterwards, we analysed the deeper (up to $m_{lim} = 28.7$ in the $g$ band) and larger ($\simeq 31' \times 33'$) Magellan/MegaCam dataset. We performed a robust analysis by carefully isolating the background galaxies, thus taking care of possible contaminations in the WL signal, which is one of the main sources of systematic uncertainties. We further verified the purity and completeness of our sample by using a reference spectroscopic redshift catalogue based on VLT/MUSE data centred on the same cluster. The reconstructed total surface mass distribution reveals the presence of three density peaks in the inner core of the cluster with significance greater than $7\sigma_k$, thus supporting the hypothesis of a complex merging phenomenon emerged from previous studies of the same cluster realised with different probes (e.g., galaxy dynamics) and performed in several bands (e.g., X-ray and radio). This picture is also confirmed by a comparison with a new high-precision SL JWST-based analysis of the central region of the cluster, with which we found a nice agreement. Indeed, not only the two surface mass distributions resulting from the two methods agree, but our cumulative total mass profile is consistent within the errors with the SL one, over the radial range where they overlap. The total mass enclosed within $2.35 \, \mathrm{Mpc}$ is $(2.56 \pm 0.26) \times 10^{15} \, \mathrm{M}_\odot$, which is consistent with previous WL analyses of the same cluster and with different probes, e.g, dynamical methods, thus making Abell 2744 one of the most massive galaxy clusters ever studied. 

\begin{acknowledgements}
The authors thank the anonymous referee for the useful comments that helped to improve the manuscript. This paper includes data gathered with the 6.5 meter Magellan Telescopes located at Las Campanas Observatory, Chile. The authors thank Amata Mercurio and Tommaso Treu for proposing us to perform an analysis of Abell 2744 with Magellan/MegaCam data to complement the SL JWST-based study by \citet{bergamini2023b}. We also thank useful discussions with Erin Sheldon about his software ngmix.
\end{acknowledgements}

\bibliographystyle{aa}
\bibliography{biblio}

\end{document}